\documentclass[aip,pop,twocolumn]{revtex4}
\usepackage{graphicx} 
\usepackage{dcolumn}
\usepackage{bm}
\usepackage{hyperref}
\usepackage[usenames,dvipsnames]{color}
\usepackage{listings}
\usepackage{wrapfig}
\usepackage{amssymb}
\usepackage{rotating}
\usepackage{amsmath,wasysym}

\begin{document}

\title{Supercriticality to subcriticality in dynamo transitions}
\author{Mahendra K. Verma}
\affiliation{Department of Physics, Indian Institute of Technology -- Kanpur 208016, India}
\author{Rakesh K. Yadav}
\altaffiliation{Previously at the Department of Physics, Indian Institute of Technology -- Kanpur 208016, India}
\email[Email : ]{yadav.r.k.87@gmail.com}
\affiliation{Max-Planck-Institut f\"ur Sonnensystemforschung, Max Planck Strasse 2, 37191 Katlenburg-Lindau, Germany}

\begin{abstract}

Evidence from numerical simulations suggest that the nature of dynamo transition changes from supercritical to subcritical as the magnetic Prandtl number is decreased. To explore this interesting crossover we first use direct numerical simulations to investigate the hysteresis zone of a subcritical Taylor-Green dynamo. We establish that a well defined boundary exists in this hysteresis region which separates dynamo states from the purely hydrodynamic solution. We then propose simple dynamo models which show similar crossover from  supercritical to subcritical dynamo transition as a function of the magnetic Prandtl number. Our models show that the change in the nature of dynamo transition is connected to the stabilizing or de-stabilizing influence of governing  non-linearities.
\end{abstract}


\maketitle

\section{Introduction}
Many natural and engineering systems exhibit transition from one state to another. Some of the prominent examples of transition are water to vapor, paramagnetic to ferromagnetic, conduction to convection (e.g., in Rayleigh B\'{e}nard convection), laminar to turbulent flow in channels, etc.  Some of these transitions are continuous, i.e., the order parameter grows smoothly from zero as the control parameter is increased, while some others exhibit a discontinuity or a finite jump in the order parameter.  The former class of transitions is called {\em supercritical}, while the latter is called {\em subcritical}. 

Magnetic field generation or dynamo process in astrophysical plasmas and in electrically conducting fluid interiors of planets and stars exhibits transition from no-dynamo or fluid state to a magnetic state~\cite{Moffat:BOOK, Brandenburg:PR2005}.  This transformation is referred to as ``dynamo transition".  Several laboratory experiments~\cite{Gailitis:PRL2001, Stieglitz:PF2001, Monchaux:PRL2007} have replicated this process in controlled laboratory experiments and report supercritical transition, whereas numerical simulation and models exhibit both supercritical~\cite{Christensen:GJI1999, Yadav:EPL2010, Krstulovic:PRE2011} and subcritical transitions~\cite{Ponty:PRL2007, Rincon:PRL2007, Reuter:NJP2009, Sahoo:PRE2010, Nigro:AJL2011, Krstulovic:PRE2011, Yadav:PRE2012}.  It is important to note that a subcritical transition is typically a combination of subcritical pitchfork bifurcation at the critical parameter value, and a saddle-node bifurcation at a lower parameter value.  A hysteresis occurs in between these two bifurcation points; in this regime, a dynamo state can exist  before the critical parameter value.   In this paper, we explain these transitions using several low-dimensional models. 

Nature of dynamo transition can play a very crucial role in our understanding of stellar and planetary dynamos.  For instance, Mars is believed to have no working dynamo, but it exhibits strong crustal magnetism.  Kuang {\it et al.}~\cite{Kuang:GRL2008} conjectured that the  sudden termination of the Martian dynamo could be due to a subcritical dynamo transition. Christensen {\it et al.}~\cite{Christensen:GJI1999} and Morin and Dormy~\cite{Morin:IJMPB2009} used numerical simulations to study dynamo mechanism in rotating spherical-shells; they observed both supercritical and subcritical dynamo transitions in such systems. Some of the important control parameters for dynamo simulations are the magnetic Prandtl number $\mathrm{Pm} = \nu/\eta$, the Reynolds number $\mathrm{Re} = U\,L/\nu$, and the magnetic Reynolds number $\mathrm{Rm} = U\,L/\eta$, where $\nu, \eta$ are the kinematic viscosity and magnetic diffusivity respectively, and $U,L$ are the large-scale velocity and length scales respectively.   Note that $\mathrm{Rm = Re\,Pm}$.  The other control parameters specific to convective dynamo are the Rayleigh number $\mathrm{Ra}$ (ratio of buoyancy and dissipation), the Ekman number (ratio of viscous force and Coriolis force), and the Roberts number (ratio of thermal diffusivity and magnetic diffusivity). 

The aforementioned spherical-shell or convective dynamos also indicate that the dynamo transition changes from supercritical to subcritical as the magnetic Prandtl number $\mathrm{Pm}$ is decreased.  Similar dependence on $\mathrm{Pm}$ has been observed in the Taylor-Green (TG) dynamo~\cite{Ponty:PRL2007, Yadav:EPL2010, Yadav:PRE2012} and shell models~\cite{Sahoo:PRE2010, Nigro:AJL2011}. Detailed numerical simulations  reveal that the critical parameter value (for instance, critical Rayleigh number) increases with the decrease of $\mathrm{Pm}$ (see e.g.~Christensen and Aubert~\cite{Christensen:GRL2006} and Ponty {\it et al.}\cite{Ponty:PRL2005}). However dynamo could occur in the subcritical regime for parameter values below the critical one as a result of a jump to the hysteresis branch.  This generic feature in observed in many dynamo models.

Low-order dynamo models, despite their simplicity, can nonetheless provide very interesting insights.  Bullard~\cite{Bullard:MPCPS1955} used a homopolar disk dynamo to study the basic dynamo mechanism, and, subsequently, Rikitake~\cite{Rikitake:MPCPS1958}  used a coupled disk dynamo model to qualitatively discern the geomagnetic reversals. Low-order models~\cite{Petrelis:JPCM2008, Gissinger:EPL2010} have also been constructed to analyze magnetic field reversals observed in the Von-Karman-Sodium (VKS) experiment~\cite{Monchaux:PRL2007, Ravelet:PRL2008}. Verma {\it et al.}~\cite{Verma:PRE2008} constructed a six-mode model to study the importance of helicity in dynamo transitions; they reported a supercritical pitchfork dynamo transition. A subcritical dynamo transition has been demonstrated by Fedotov {\it et al.}~\cite{Fedotov:PRE2006} in a perturbative $\alpha\Omega$-dynamo model. Krstulovic {\it et al.}~\cite{Krstulovic:PRE2011} also briefly discused a model containing two nonlinear equations to explain the $\mathrm{Pm}$ dependence of the supercritical and subcritical dynamo transitions. Weiss~\cite{Weiss:GAFD2010} provides a detailed review of the important low-order dynamo models which have been used to understand planetary and solar dynamos. 

In this paper we discuss some of the new properties of the Taylor-Green dynamo related to the subcriticality.  For example, the final state (dynamo or no-dynamo) of a system depends quite critically on the initial condition, which are called ``basins of attraction".  We then describe a pedagogical one-dimensional model and construct 3-mode models that capture both supercritical and subcritical dynamo transitions.  Our models show a  good qualitative agreement with the recently performed numerical simulations of the TG dynamo and the convective dynamos~\cite{Morin:IJMPB2009, Yadav:EPL2010, Krstulovic:PRE2011, Yadav:PRE2012}. 

The article is structured as follows: Section~\ref{sec:TG_DNS} contains a direct numerical simulation (DNS) study of a TG dynamo exhibiting  subcritical dynamo transition.   In Sec.~\ref{sec:1D}, we discuss one-dimensional equations which exhibit supercritical and subcritical bifurcations.  We construct 3-mode models for the TG dynamo and the convective dynamo in Sections~\ref{sec:3mode} and~\ref{sec:convective_dynamo} respectively.   We conclude in Sec.~\ref{sec:conclusions}.  In Appendix~\ref{sec:nlin} we provide a derivation for the 3-mode model.   Appendix~\ref{sec:isola} contains a brief description on the ``isola" bifurcation exhibited by the one-dimensional model.  

\section{\label{sec:TG_DNS}Subcritical Dynamo Transition in Taylor-Green Dynamo} 

Many researchers~\cite{Nore:PP1997, Mininni:APJ2005, Ponty:PRL2005, Ponty:PRL2007, Dubrulle:NJP2007, Yadav:EPL2010,  Yadav:PRE2012} have explored the TG dynamo using numerical simulations. Recently,   a supercritical dynamo transition was observed for $\mathrm{Pm}$ unity \cite{Yadav:EPL2010}, and a subcritical dynamo transition and a hysteresis cycle was observed for $\mathrm{Pm}$ of half~\cite{Yadav:PRE2012}.  To further probe the subtle nature of hysteresis, we perform a set of numerical simulations of the TG dynamo for  $\mathrm{Pm}=1/2$.

The fluid velocity $\mathbf u$ and the magnetic field $\mathbf B$ in a dynamo mechanism are governed by  the magnetohydrodynamic (MHD) equations:
\begin{eqnarray}
\partial_{t}\mathbf{u}+ (\mathbf{u} \cdot \nabla) \mathbf{u} & = &
-\nabla p+ (\mathbf{J} \times \mathbf{B} ) + \nu\nabla^{2}\mathbf{u}+\mathbf{F}, \label{eq:MHD_vel}\\
\partial_{t}\mathbf{B} & = &
\nabla \times ( \mathbf{u} \times \mathbf {B} )+\eta\nabla^{2}\mathbf{B}, \\
\nabla \cdot \mathbf{u} & = & 0,   \\
\nabla \cdot \mathbf{B} & = & 0,  \label{eq:div_B_0}
\end{eqnarray}
where $\mathbf J$, $\mathbf F$, $p$, $\nu$, and $\eta$ represent the current density, external driving forcing, hydrodynamic pressure, kinematic viscosity, and magnetic diffusivity, respectively.  Note that ${\bf J} = \nabla \times {\bf B}$.  The density of the fluid is chosen to be unity.  In our simulation we employ TG forcing, which is defined as
\begin{eqnarray}
\mathbf{F}(k_0) & = & F_0
\left[
\begin{array}{c}
\sin(k_0 x)\cos(k_0 y)\cos(k_0 z) \\
-\cos(k_0 x)\sin(k_0 y)\cos(k_0 z) \\
0
\end{array}
\right], 
\end{eqnarray}
to study the dynamo transition.  Here, $F_0$ is the forcing amplitude and $k_0$ (=2) defines the length scale of the forcing.   We use a pseudospectral code Tarang~\cite{Verma:arxiv2012} to numerically solve Eqs.~(\ref{eq:MHD_vel}-\ref{eq:div_B_0}) with $\mathrm{Pm} = 1/2$ ($\nu=0.1$ and $\eta=0.2$) in a cube of dimensions $(2 \pi)^3$ having periodic boundary conditions on all sides. We discretize the simulation box into  $64^3$ grids. Interested reader is referred to Ref.~\onlinecite{Yadav:PRE2012} for more details about the numerical procedure.

\begin{figure}
\begin{center}
\includegraphics[scale=0.55]{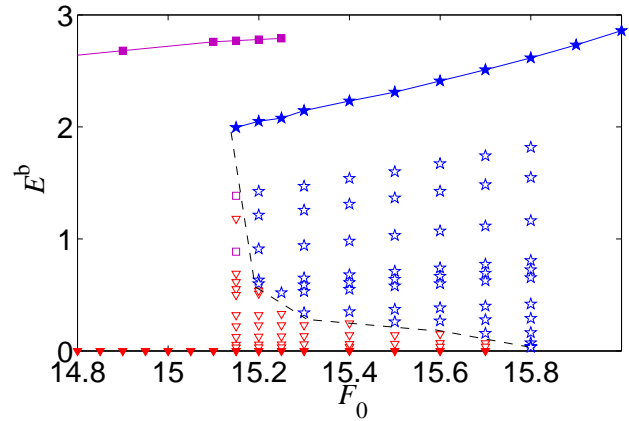}
\caption{Bifurcation diagram constructed using various dynamo states of the TG dynamo for $\mathrm{Pm}=1/2$. The filled blue-stars connected by solid line represent a stable dynamo branch. The runs with hollow blue-stars as initial conditions go to the  dynamo branch (filled blue-stars), while the runs with hollow red-triangles as initial conditions settle down to the no-dynamo states (filled red-triangles). The initial conditions denoted by hollow purple-squares near the hysteresis boundary stabilize to a different dynamo branch represented by filled purple-squares.}
\label{fig:subcritical_TG_DNS}
\end{center}
\end{figure}

Figure~\ref{fig:subcritical_TG_DNS} contains results obtained using DNS. As reported by Yadav {\it et al.}~\cite{Yadav:PRE2012}, at $F_0 = 15.8$ we first observe a fixed-point dynamo state (temporally non-fluctuating); here, the magnetic energy $E^b$ shows a finite jump to approximately $2.6$. However, when the dynamo state at $F_0 = 15.8$ is used as an initial condition and the forcing amplitude is gradually decreased, the dynamo state continues along the trail (filled blue-stars) till $F_0 \approx15.15$, at which point there is a sudden jump to a fluid state (filled red-triangles). Thus, $F_0=15.15:15.8$ is a hysteresis band for $\mathrm{Pm}=1/2$ TG dynamo.  From our simulations we can deduce a subcritical pitchfork bifurcation at $F_0=15.8$ and saddle-node bifurcation at $F_0=15.2$.  

The final state of the hysteresis zone depends quite critically on the initial condition.  To explore the basins of attraction of the $E^b=0$ and the $E^b > 0$ states, we construct a set of initial conditions using the dynamo states on the filled blue-star trail in Fig.~\ref{fig:subcritical_TG_DNS}. For the initial conditions, we keep the velocity field same as that of the dynamo state, but the magnetic field is quenched by a factor, e.g., 2, 4, 6, etc.  The set of initial conditions used in this work are indicated by hollow data points in Fig.~\ref{fig:subcritical_TG_DNS}.  Interestingly, for a given $F_0$, all the runs with hollow blue-stars as initial conditions stabilize to the corresponding dynamo states (filled blue-stars). On the contrary, the runs with hollow red-triangles as initial conditions settle to the corresponding no-dynamo states (filled red-triangles). Thus, in the hysteresis zone, the regions with hollow blue-stars and hollow red-triangles form the basin of attraction of the dynamo and the no-dynamo states respectively.  The dashed black curve separating the two regions is the unstable manifold; this unstable manifold may be a fractal~\cite{Sahoo:PRE2010}, whose computation requires a large number of runs with different initial conditions near the boundary.   These results demonstrate  a critical dependence of the final state on the initial condition,  a result that has  important implications on the numerical simulations and experiments of dynamo. Earlier, Childress and Soward~\cite{Childress:PRL1972} discussed the influence of initial condition in the context of weak and strong field dynamos.

We observe another interesting feature that the runs with the  initial conditions denoted by hollow purple-squares stabilize to the corresponding filled purple-square dynamo state (at $F_0 = 15.15$), which belongs to a different dynamo branch.  This new branch originates from the edge of the hysteresis, and it is related to the high-dimensionality of the system.  The other dynamo states on the purple trail were constructed using the filled purple-square dynamo state for $F_0 = 15.15$. The above analysis demonstrates the complex nature of the dynamo transition, especially in the hysteresis zone.    

The dynamo transition in the TG dynamo has certain similarities with the convective dynamo.  Morin and Dormy~\cite{Morin:IJMPB2009} showed that in a spherical-shell dynamo, the transition is supercritical, subcritical, and isola as the magnetic Prandtl number is decreased.  Sreenivasan and Jones~\cite{Sreenivasan:JFM2011} investigated the subcriticality of dynamo transition in spherical-shell dynamos in greater detail.   The TG and convective dynamos  have many degrees of freedom.  However, a common feature in these systems is that  the large-scale modes dominate near the dynamo transition.   In the next three sections we will describe several low-dimensional models based on large-scale modes that reproduce some of the aforementioned features, thus enhancing our understanding of the dynamo transition.

\section{\label{sec:1D}One-dimensional Model} 
In this section we describe a popular pedagogical one-dimensional model that exhibits  supercritical and subcritical transitions~\cite{Strogatz:BOOK, Bhattacharjee:BOOK}. A simpler version of the model which exhibits supercritical bifurcation is
\begin{eqnarray}
\dot{X}={C_1}X+{C_3}X^3.  \label{eq:supercritical}
\end{eqnarray}
where $X$ could represent the large-scale magnetic field.  The parameters $C_1$ and $C_3$ are functions of the system's parameters, e.g., amplitude  of the external force and the $\mathrm{Pm}$.  The cubic term $C_3 X^3$ represents an effective nonlinear interaction among the large-scale modes.   

The model attains a fixed-point solution $X = X^*$ asymptotically (i.e. as $t \rightarrow \infty$), except when $C_1, C_3 >0$ for which $X\rightarrow +\infty$ or $-\infty$ depending on the initial condition.  For negative values of $C_3$, the solution $X^*=0$ is stable for $C_1<0$.  For $C_1 > 0$, $X^*=0$ becomes unstable, and two new stable solutions $\pm \sqrt{-C_1/C_3}$ are born.  The system picks one of them depending on the initial condition.   

A  generalization of the above model that exhibits both supercritical and subcritical transitions is~\cite{Strogatz:BOOK, Bhattacharjee:BOOK}
\begin{eqnarray}
\dot{X}={C_1}X+{C_3}X^3+{C_5}X^5.  \label{eq:combined}
\end{eqnarray}
The new ${C_5}X^5$ term is a higher-order nonlinear term.
The fixed-point solutions of the above equation are
\begin{equation}
X^* = 0, \hspace{0.25cm} 
\label{eq:soln0}
\end{equation}
\begin{eqnarray}
X_{++}^*  & = &   \left( \frac{-C_3 + \sqrt{C_3^2 - 4 C_1 C_5}}{2 C_5} \right)^{1/2}, \\
X_{+-}^*  & = &   \left( \frac{-C_3 - \sqrt{C_3^2 - 4 C_1 C_5}}{2 C_5} \right)^{1/2}, \\
X_{-+}^*  & = & - \left( \frac{-C_3 + \sqrt{C_3^2 - 4 C_1 C_5}}{2 C_5} \right)^{1/2}, \\
X_{--}^*  & = & - \left( \frac{-C_3 - \sqrt{C_3^2 - 4 C_1 C_5}}{2 C_5} \right)^{1/2}.
\label{eq:solnX}
\end{eqnarray}
The nature of these fixed-point solutions depends on the values of the parameters $C_1, C_3$ and $C_5$.  

\subsection{Supercritical Transition}
When both $C_3$ and $C_5$ are negative, we obtain a supercritical transition at $C_1=0$, as shown in Fig.~\ref{fig:1d}(a) (here $C_3 = C_5=-1$).  $X^*=0$ is the only stable solution for $C_1 < 0$. However, for $C_1>0$,  $X^*=0$ becomes unstable, and $X^*_{+-}$ and $X^*_{--}$ are the only possible stable solutions;  other solutions $X^*_{-+}$ and $X^*_{++}$  become complex.   

\subsection{Subcritical Transition}
\begin{figure}
\begin{center}
\includegraphics[scale=0.40]{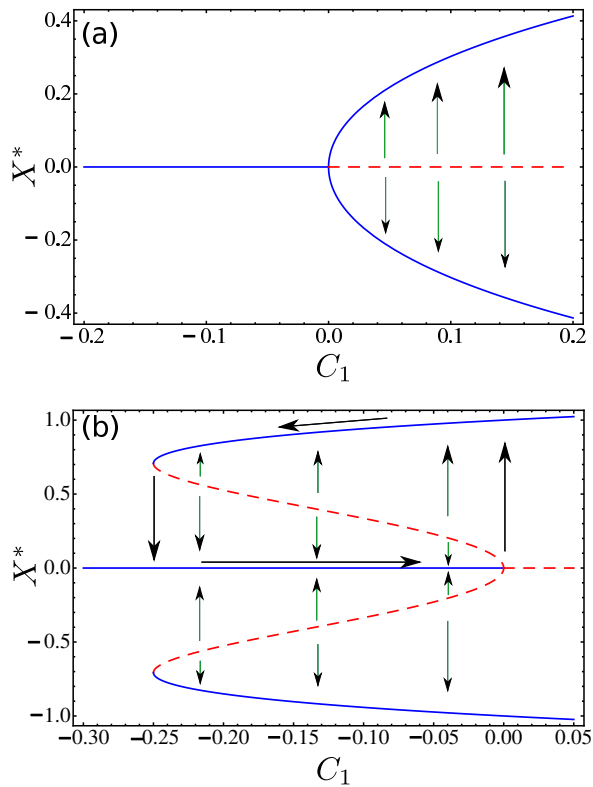}
\caption{Bifurcation diagrams for Eq.~(\ref{eq:combined}) portraying (a) supercritical transition for $C_3 = C_5=-1$, and (b) subcritical transition $C_3 =  1$ and $C_5 = -1$.  The blue solid-curve and the red dashed-curve are stable and unstable branches respectively. The light green arrows represent the ``flow" directions of the initial conditions in respective regions. In figure (b), the black arrows depict the hysteresis cycle of the subcritical transition. }
\label{fig:1d}
\end{center}
\end{figure}

A subcritical transition is observed when $C_3>0$ and $C_5<0$.  The $C_5 X^5 $ term stabilizes the system for large $X$.  The bifurcation diagram for a generic case $C_3=1$ and $C_5=-1$ is exhibited in Fig.~\ref{fig:1d}(b), which illustrates a  subcritical pitchfork bifurcation at $C_1 =0$ and a saddle-node bifurcation at $C_1 = C_3^2/(4 C_5)$~\cite{Strogatz:BOOK, Bhattacharjee:BOOK}.   Note that all fixed-point solutions $X^*_{\pm \pm}$ are real for $C_3^2/(4 C_5)\leq C_1 \leq 0$, with $X^*_{+-}$ and   $X^*_{--}$ being stable, and $X^*_{++}$ and $X^*_{-+}$ being unstable. The stable solutions $X^*_{+-}$ and $X^*_{--}$ continue to exist for $C_1 > 0$.

The hysteresis  cycle represented by the black arrows in Fig.~\ref{fig:1d}(b) results due to the aforementioned bifurcations.  For $C_3^2/(4 C_5)<C_1<0$, initial conditions below the unstable manifold (red dashed-curve) settle down to $X^*=0$, while the ones above the unstable manifold settle down to $X^* > 0$ solutions (blue solid-curve), as shown in Fig.~\ref{fig:1d}(b).  Thus, initial condition plays a critical role in determining the final state of a subcritical system.

Equation~(\ref{eq:combined}), which has one more nonlinear terms (the 5th order) than Eq.~(\ref{eq:supercritical}), exhibits both supercritical and subcritical transition depending on the sign of $C_3$.  In the supercritical transition $C_3<0$, and hence the term $C_3 X^3$ saturates the solution.   However, for the subcritical transition $C_3>0$, leading to a  rapid growth of the solution which is further saturated by the $C_5 X^5$ term.   We will show later in the paper that a variation of the magnetic Prandtl number could change the sign of $C_3$, thus inducing a transition from supercritical to subcritical transition.

The above one-dimensional model captures the salient features of supercritical and subcritical dynamo transitions exhibited by TG and convective dynamos.  Yet, the role of $\mathrm{Pm}$ and other parameters is not explicit in this model because all the effects of different nonlinearities have been clubbed into the $X^3$ and $X^5$ terms of the model.  To overcome this deficiency, at least partially, we introduce three large-scale modes, one velocity and two magnetic, and write a minimal model which could capture the salient features of the dynamo transition.  The two magnetic modes could correspond to the dipolar and the quadrupolar modes of the magnetic field in convective dynamo~\cite{Christensen:GJI1999, Morin:IJMPB2009}, or to the dominant magnetic Fourier modes of the TG dynamo~\cite{Yadav:EPL2010, Yadav:PRE2012}. 

\section{\label{sec:3mode}Three-mode  Model for the Taylor-Green Dynamo} 

Numerical simulations provide valuable information about the dynamo transition.  Yet, huge datasets produced by large-scale simulations hide the essential physics to some degree.  To probe the dynamics of the dynamo transition, we construct a low-dimensional model consisting of one velocity Fourier mode $U$, and two magnetic Fourier modes $B_1$ and $B_2$. In the TG dynamo, numerical simulations reveal that the modes $\mathbf{u}(2,2,2)$, $\mathbf{B}(0,0,1)$, and $\mathbf{B}(-2,-2,-1)$ are the most energetic and dominate of the system~\cite{Yadav:EPL2010, Yadav:PRE2012}; these modes are used for constructing the model.  Interestingly these modes also form a nonlinear triad (${\mathbf k = \mathbf  p+ \mathbf  q}$).  

In  Appendix~\ref{sec:nlin} we derive a low-dimensional model using Galerkin truncation and by making certain assumptions.  The  equations of our model in nondimensional form are
\begin{eqnarray}
\dot{U} & = & f- U-(\alpha+1)B_{1}B_{2}, \label{eq:up_dot}\\ 
\dot{B}_{1} & = & - \frac{1}{\mathrm{Pm}} B_{1} + \alpha U B_{2} -\beta B_2^2 B_1, \label{eq:B1p_dot}\\
\dot{B}_{2} & = & - \frac{1}{\mathrm{Pm}}  B_{2}+U B_{1} +\beta B_1^2 B_2, \label{eq:B2p_dot}
\end{eqnarray}
where $\alpha$ and $\beta$ are constants.   Note that the quadratic ($B_1 B_2, U B_2, U B_1$) and the cubic ($B_2^2 B_1, B_1^2 B_2$) nonlinearities conserve the total energy ($U^2+B_1^2+B_2^2$) of the system.  Also, the model preserves the $\mathbf u \rightarrow \mathbf u$ and  $\mathbf B \rightarrow -\mathbf B$ symmetry of the MHD equations.

The above model shows a dynamo transition, with constant magnetic field appearing near the onset.  To understand the system behaviour near the transition, we compute the  fixed point solution of the above set of equations.    The nonmagnetic or fluid solution $B_1=B_2=0$ and $U=f$ is a trivial solution of the model.  Finding the dynamo  solution however is more involved, and is obtained by first writing $U$ and $B_2$ as (from Eqs.~(\ref{eq:up_dot}) and (\ref{eq:B2p_dot})): 
\begin{eqnarray}
U & = & f-(\alpha+1) B_1 B_2   \label{eq:U_approx}  \\
B_2 & = & \frac{\mathrm{Pm} f B_1}{1+\mathrm{Pm} (\alpha-\beta+1) B_1^2}.  \label{eq:B2_approx} 
\end{eqnarray}
After this, we substitute the above expressions of $U$ and $B_2$ in Eq.~(\ref{eq:B1p_dot}), which yields 
\begin{eqnarray}
& &\underbrace{ \left[-1+\alpha f^2 \mathrm{ Pm}^2   \right]}_{C_{1}}B_{1} \nonumber \\
& + &\underbrace{  \mathrm{Pm}  \left[  2(\beta-\alpha-1) - \beta(\alpha+1) f^2 \mathrm{Pm}^2 \right]}_{C_{3}}B_{1}^{3}  \nonumber \\ 
& + &  \underbrace{ \left[ - (\alpha-\beta+1)^2 \mathrm{Pm}^2  \right]}_{C_{5}}B_{1}^{5} = 0. \label{eq:B1_approx} 
\end{eqnarray}
The Eq.~(\ref{eq:B1_approx}) is a fifth order algebraic equation whose roots are $B_1=0$ (trivial), and four others (real or complex).  The equation is reminiscent of Eq.~(\ref{eq:combined}) that exhibits subcriticality. Using Eqs.~(\ref{eq:U_approx}-\ref{eq:B1_approx}) we can conclude that  that all the three modes $U, B_1$ and $B_2$ are nonzero for the dynamo branch in both supercritical and subcritical cases.

We now explore the possibility of subcritical and supercritical behaviour in Eq.~(\ref{eq:up_dot})-(\ref{eq:B2p_dot}) by making an analogy with the 1D model. The dynamo transition forcing $f=f_{c1}$ is obtained by setting $C_1=0$ (of Eq.~(\ref{eq:B1_approx})), which yields
\begin{equation}
|f_{c1}| = \frac{1}{\mathrm{Pm}\sqrt{\alpha}}.
\end{equation}
Supercritical to subcritical transition takes place when $C_3=0$, or at $f=f_{c2}$ with 
\begin{equation}
|f_{c2}| = \frac{1}{\mathrm{Pm}}  \sqrt{\frac{2(\beta-\alpha-1)}{ \beta (\alpha+1)}}.
\end{equation}
Note that $C_3< 0$ for $|f|>|f_{c2}|$, and vice versa.  Hence, the dynamo transition is supercritical for $|f_{c2}| <  |f_{c1}|$ or $\beta > \beta_c = 2\alpha(\alpha+1)/(\alpha-1)$, and subcritical  for the reversed condition.  

To illustrate the above transition, we choose $\alpha=2$, which yields $\beta_c= 12$.  Therefore, $\beta<\beta_c=12$ yields supercritical  transition, and $\beta>12$ yields subcritical  transition.  However,  instead of changing $\beta$, we model $\beta=\beta_c /\mathrm{Pm}$ and vary $\mathrm{Pm}$.  We obtain supercritical transition for $\mathrm{Pm} >1$, and subcritical for $\mathrm{Pm} <1$.  We illustrate these transitions using $\mathrm{Pm}=2$ (supercritical) and 1/2 (subcritical).

Furthermore, using Eq.~(\ref{eq:B1_approx}) we can construct the following potential function for $B_1$: 
\begin{eqnarray}
V(B_1) & = & -\int (C_1 B_1 + C_3 B_3^3 + C_5 B_1^5)dB_{1} \nonumber \\
& = & -C_1\frac{B_{1}^{2}}{2}-C_3\frac{B_{1}^{4}}{4}-C_5\frac{B_{1}^{6}}{6}. 
\end{eqnarray}
This function provides an elegant way of visualizing the nature of the dynamo states~\cite{Strogatz:BOOK, Bhattacharjee:BOOK}.  In this description, the stable and unstable dynamo states are represented by the ``valleys" and ``hills" of the potential function, respectively.  The potential function also provides information about the dynamics of the system.   Given an initial condition, the slope of the potential provides the local direction of motion.

\subsection{Supercritical Transition} \label{supercritical}
\begin{figure}
\vspace{20pt}
\begin{center}
\includegraphics[scale=0.6]{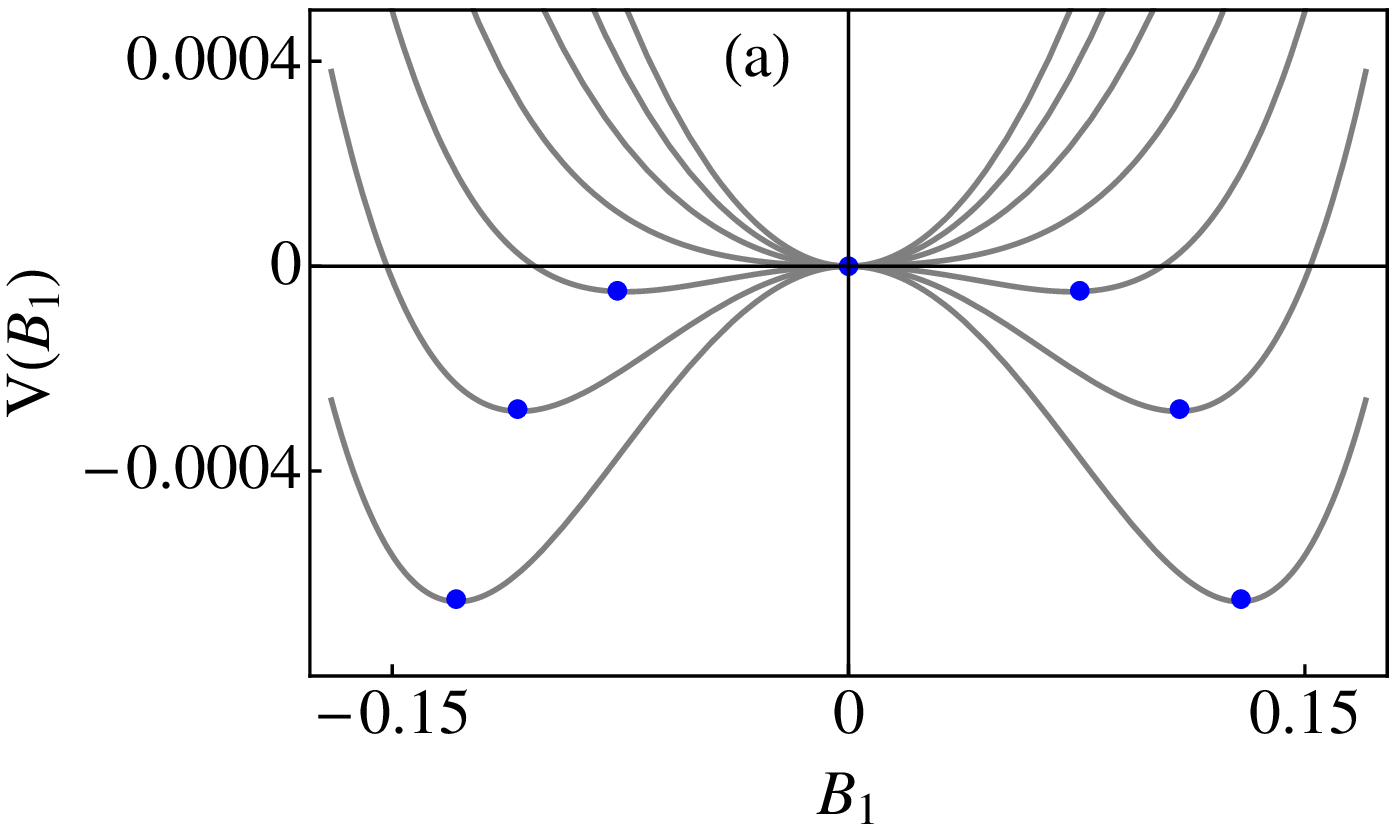} \\
\includegraphics[scale=0.6]{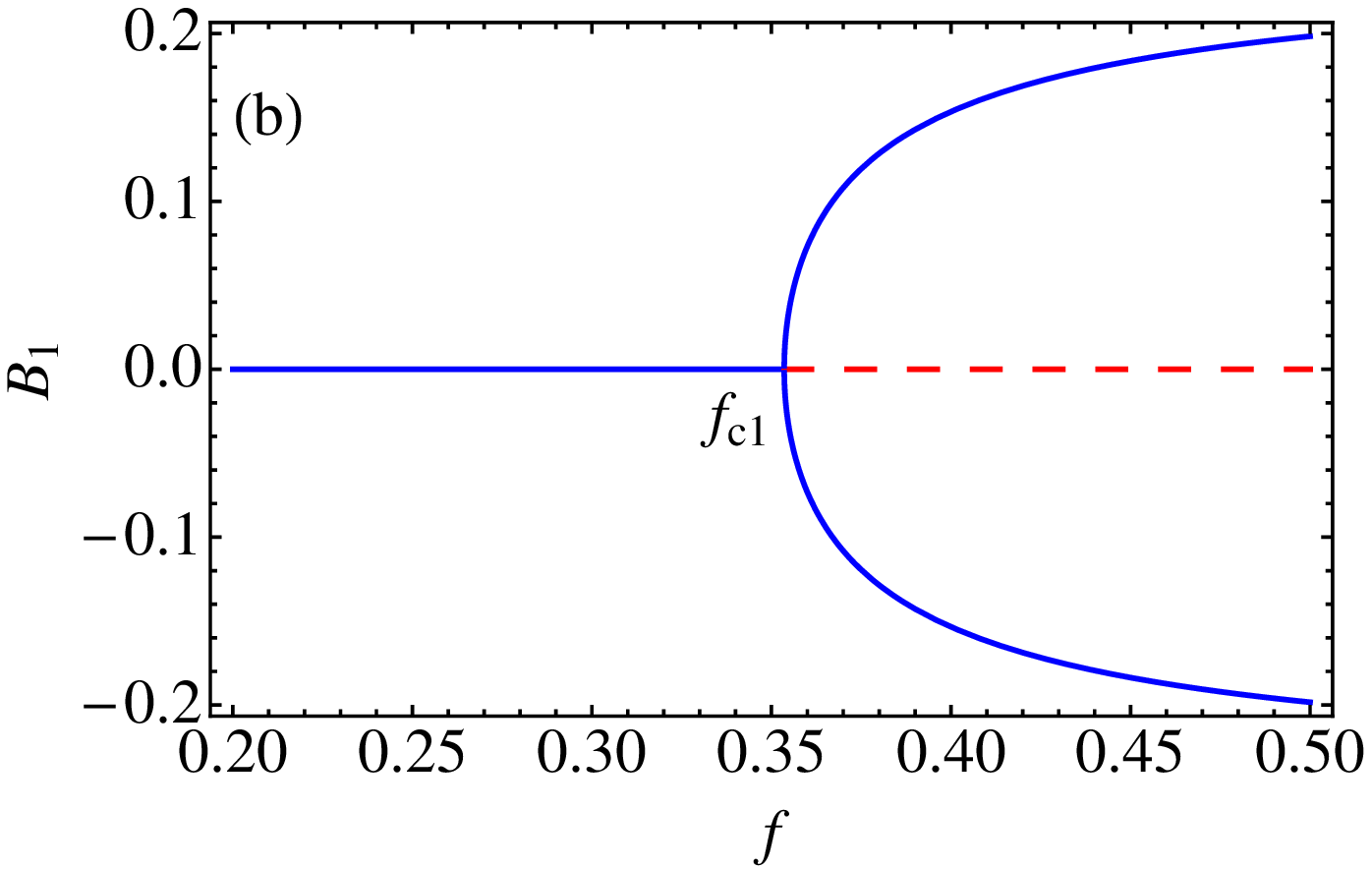}
\caption{For the 3-mode model for the TG dynamo with $\alpha=2$, $\mathrm{Pm}=2$, and $\beta=6$:  (a) Plot of the potential function, and (b) a bifurcation diagram illustrating  supercritical transition.  In the top panel, the forcing $f$ is varied from 0.32 to 0.38 in steps of 0.01.  In (a), the filled-circles mark the stable valleys, and in (b), solid-blue (dashed-red) curves represent the dynamically stable (unstable) branches.
}
\label{fig:TG_supercritical}
\end{center}
\end{figure}

Here we present a representative example of a supercritical dynamo by taking $\alpha=2$, $\mathrm{Pm} =2$, and $\beta=6$.  The dynamo states are born when $C_1 > 0$, i.e. when
\begin{equation}
f> f_{c1} = \frac{1}{\mathrm{Pm} \sqrt{\alpha} } = \frac{1}{2\sqrt{2}}. \nonumber
\end{equation}
The potential function $V(B_1)$ and a bifurcation diagram for the above parameters are illustrated in Fig.~\ref{fig:TG_supercritical}. The valleys at $B_1=0$ and $B_1\ne0$ represent no-dynamo and dynamo states respectively (blue dots in Fig.~\ref{fig:TG_supercritical}(a)).

\subsection{\label{sec:subcritical}Subcritical Transition} 
\begin{figure}
\vspace{20pt}
\begin{center}
\includegraphics[scale=0.6]{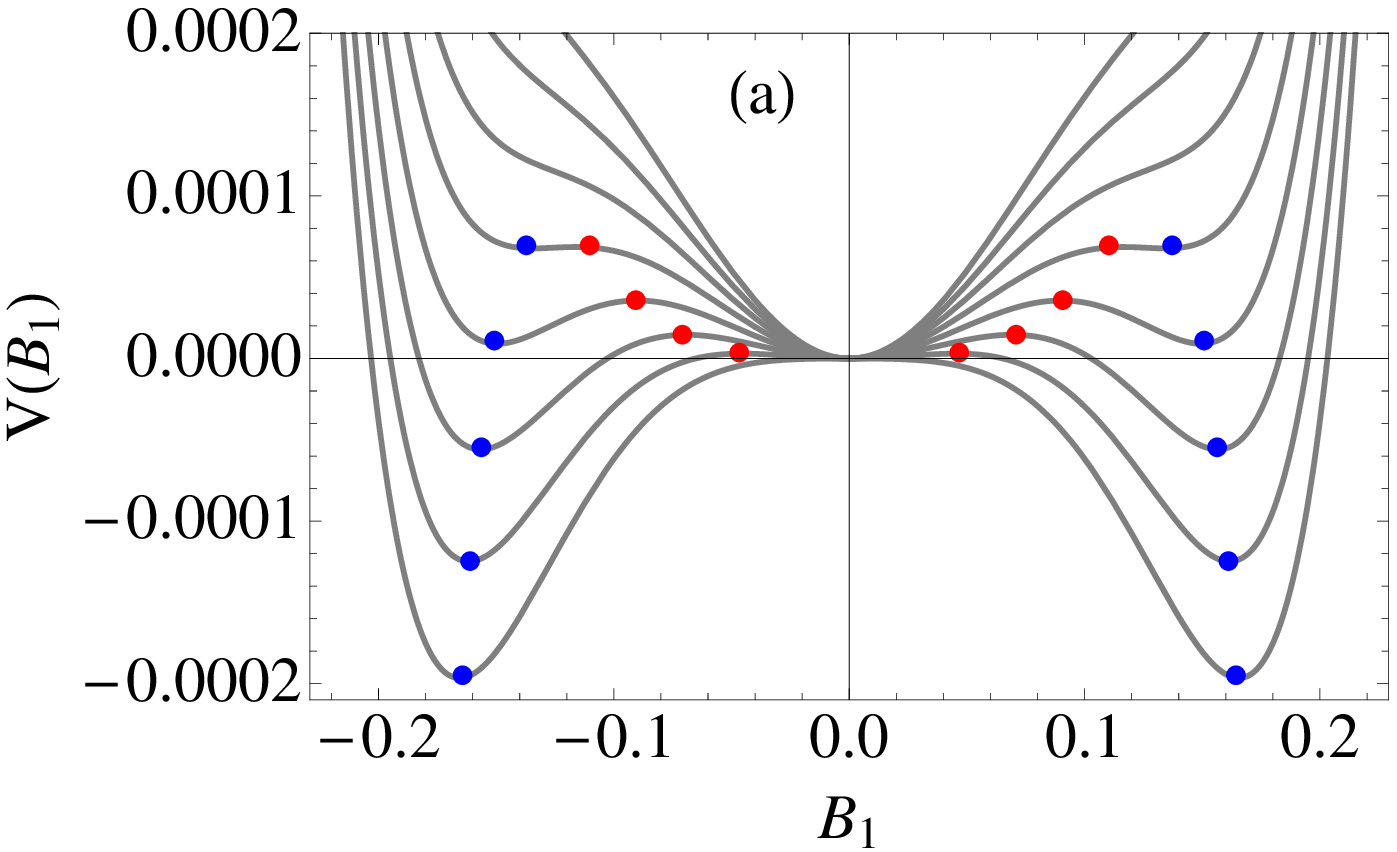} \\
\includegraphics[scale=0.6]{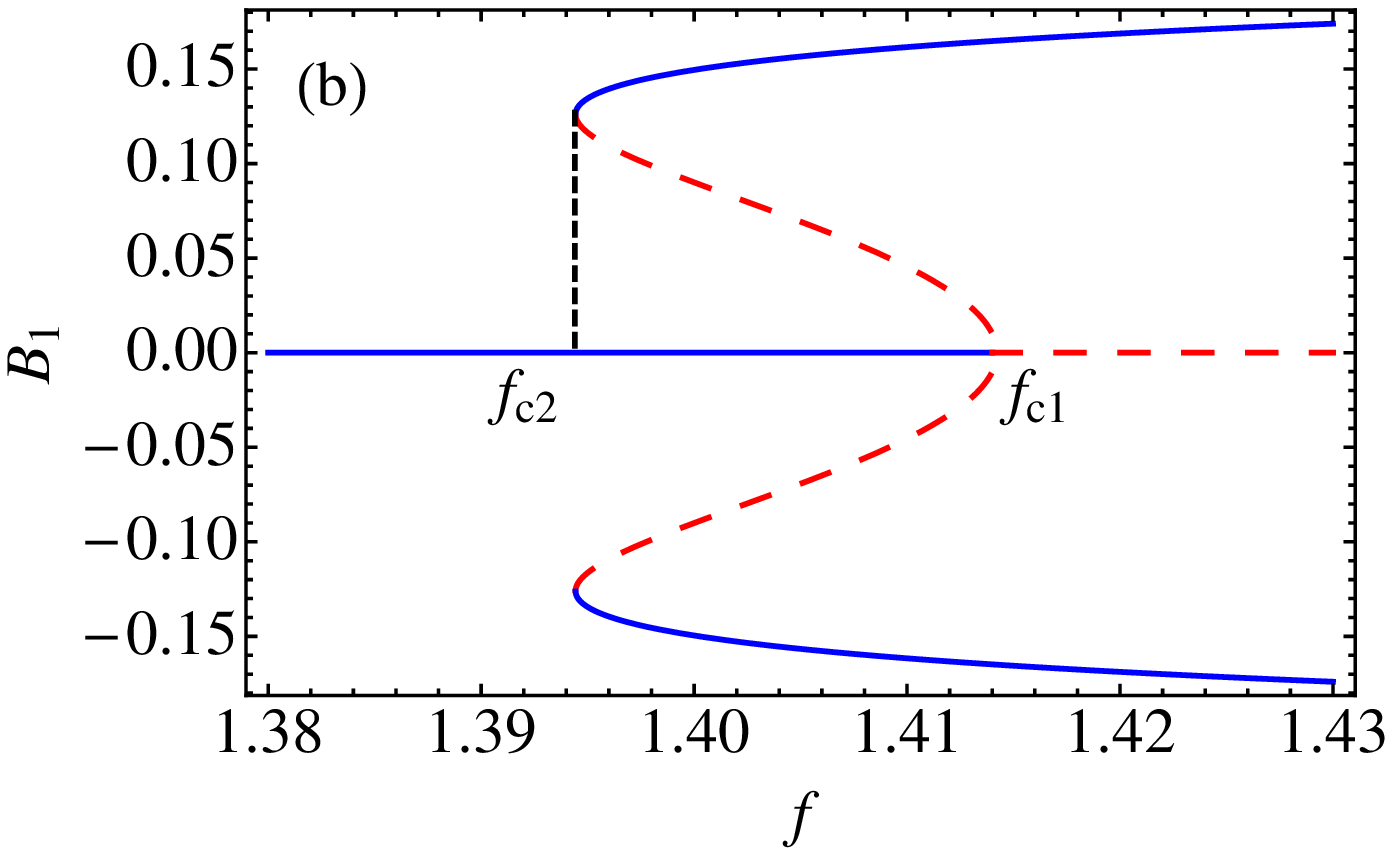}
\caption{For the 3-mode model for the TG dynamo with $\alpha=2$, $\mathrm{Pm}=1/2$, and $\beta=24$:  (a) Plot of the potential function, and (b) a bifurcation diagram illustrating  supercritical transition.  In the top panel, the forcing $f$ is varied from 1.38 to 1.415 in steps of 0.005.   Here the blue (red) circles represent the stable (unstable) dynamo states corresponding to the solid (dashed) lines of (b).
}
\label{fig:subcritical}
\end{center}
\end{figure}

To illustrate a subcritical dynamo transition, we choose  $\alpha=2$, $\mathrm{Pm}=1/2$, and $\beta=24$. For these parameters,  $C_3 >0$ and $C_5 <0$, thus satisfying the condition for subcritical dynamo transition.   In Fig.~\ref{fig:subcritical} we plot the potential functions and the bifurcation diagram near $f=f_{c1} = \sqrt{2}$. At $f=f_{c1}$, $B_1$ jumps from zero to a finite value (see Fig.~\ref{fig:subcritical}(b)).  As we decrease $f$ from $f_{c1}$ with the above dynamo state as an initial condition, the system continues to have nonzero $B_1$ until $f=f_{c2}\approx 1.394$, after which it attains $B_1=0$ state.  

In Fig.~\ref{fig:subcritical}(a) we plot the potential functions for $f$ in range $[1.38:1.415]$.  In this figure,  we observe a single valley (stable state) at $B_1=0$  for $f\le f_{c2}$, and two valleys at $\pm B_1$ and a hill at $B_1=0$ for $f \ge f_{c1}$.  For $f_{c2} \le f \le f_{c1}$, the system has three valleys and two hills.  The system can ``roll" to one of the valleys depending on the initial condition, which explains the sensitivity of the final state on the choice of an initial condition.

\begin{figure}
\vspace{20pt}
\begin{center}
\includegraphics[scale=0.6]{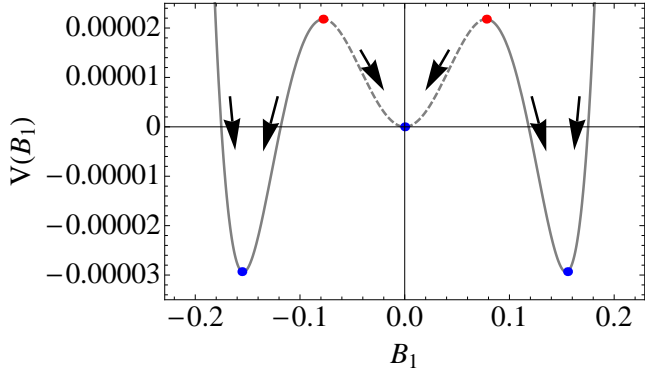}
\caption{A plot of the potential function for the 3-mode model with $\alpha=2$, $\mathrm{Pm}=1/2$, $\beta=24$, and $f=1.403$. Note that $f_{c2} < f < f_{c1}$.  The dashed curve represents the initial conditions for the no-dynamo state (filled blue-dot at $B_1=0$), while the solid curve represents the initial conditions for the dynamo state (filled blue-dots at $B_1\ne0$). The arrows represent the rolling direction of the dynamical system.
}
\label{fig:single_potential}
\end{center}
\end{figure}

To better understand the role of initial condition on the final state of the system, in Fig.~\ref{fig:single_potential} we plot a potential function for $f=1.403$. This forcing value lies in between $f_{c2}$ and $f_{c1}$, thus in the hysteresis  zone. If the initial condition lies on the dashed region of the potential function, then the system  rolls down to the $B_1=0$ valley (no-dynamo state). On the other hand, if the initial condition lies on the solid regions of the potential function, then the system rolls down to a dynamo state (corresponding to the $B_1\ne 0$ valleys), as indicated by the arrows. This behavior is possible only for $f_{c2} < f < f_{c1}$. For  $f>f_{c1}$, the two $B_1\ne0$ unstable hills merge with $B_1=0$ state.

The above results indicate that the 3-mode dynamo model captures the supercritical to subcritical dynamo transition nicely.  Note that the nature of transition changes from supercritical to subcritical depending on the sign of $C_3$, which is related to the variation of $\mathrm{Pm}$ in our model.  It also exhibits a sensitive dependence of the final states on the initial conditions.   Thus, the 3-mode model provides further  insights into the dynamo transition than the 1D model.

In the next section we construct another 3-mode model suitable for convective dynamos and test whether it could explain the dynamo transition. 

\section{\label{sec:convective_dynamo}Convective Dynamo} 
Our 3-mode model for the convective dynamo has the same structure as that for the TG dynamo, except for the forcing term.  The magnetic modes $B_1$ and $B_2$ could be interpreted as the dipolar and the quadrupolar magnetic fields respectively, and $U$ as the large-scale velocity mode.  The convective dynamo is driven by the buoyancy, hence we model the combined effects of the forcing and the dissipative terms as $r U - U^3$ with $r = \mathrm{Ra}-\mathrm{Ra}_c$, where  $\mathrm{Ra}$ and $\mathrm{Ra}_c$ are the Rayleigh number and critical Rayleigh number respectively.  Clearly, a nonzero $U$ would result only for $r = \mathrm{Ra} - \mathrm{Ra}_c > 0$ (with $B=0$), consistent with the convective instability.  This forcing is a reasonable model near the dynamo transition.  We assume the same form of nonlinearity as the TG dynamo and also ignore rotation in our model.   Under these assumptions, a 3-mode model for the convective dynamo near the dynamo transition is
\begin{eqnarray}
\dot{U} & = & r U - U^3 -(\alpha+1)B_{1}B_{2}, \label{eq:CD_up_dot}\\ 
\dot{B}_{1} & = & - \frac{1}{\mathrm{Pm}} B_{1} + \alpha U B_{2} -\beta B_2^2 B_1, \label{eq:CD_B1p_dot}\\
\dot{B}_{2} & = & - \frac{1}{\mathrm{Pm}}  B_{2}+U B_{1} + \beta B_{1}^{2} B_2. \label{eq:CD_B2p_dot}
\end{eqnarray}

We solve for the fixed points of the above model following the same procedure as described in the previous section.  The pure fluid solution is the trivial solution: $U=\pm\sqrt{r}$ and $B_1 = B_2=0$.  We also obtain the following dynamo solution after several algebraic manipulations:
\begin{eqnarray}
 & &\underbrace{\left[ r \alpha -\frac{1}{\mathrm{Pm}^2} \right]}_{C_{1}}B_{1} \nonumber \\ 
& + & \underbrace{  \mathrm{Pm}\left[  \frac{3 \beta}{\mathrm{Pm}^2} - (2\alpha+1)r \beta - \alpha(\alpha+1) \right]}_{C_{3}}B_{1}^{3}  \nonumber \\ 
& + & \underbrace{ \left[  -3\beta^2 + (\alpha+1)(r\beta+\alpha+1) \beta\mathrm{Pm}^2 \right]}_{C_{5}}B_{1}^{5} \nonumber \\ 
& - &\beta^3 \mathrm{Pm}  B_1^7= 0, \label{eq:CD_B1_approx}  \\
U & = & \pm \sqrt{r - \frac{(\alpha+1)\mathrm{Pm} B_1^2}{1-\beta \mathrm{Pm} B_1^2}} \\
B_2 & = & \frac{\mathrm{Pm} U B_1}{(1-\beta \mathrm{Pm} B_1^2)}
\end{eqnarray}
which has a similar form as the fixed-point solution of the 3-mode model of the TG dynamo, except for the $B_1^7$ term.  We neglect this highest order term since, near dynamo transition where $B_1$ is small, it would be very small compared to the $B_1^3$ and $B_1^5$ terms.

Dynamo transition ($|B_1|,|B_2|>0$) takes place when $C_1$ changes sign from negative to positive, i.e., when 
\begin{equation}
r > r_c = \frac{1}{\alpha \mathrm{Pm}^2}.
\end{equation}
The pure fluid solution is a stable solution for $r<r_c$.  As shown is Section~\ref{sec:1D}, the transition is supercritical when $C_3<0$ at the transition, and subcritical when $C_3>0$ and $C_5<0$.  Therefore we compute $C_3(r = r_c)$ and test its sign, i.e.,
\begin{eqnarray}
C_3(r = r_c)  & = &  \frac{3 \beta}{\mathrm{Pm}^2} - (2\alpha+1) \beta \frac{1}{\alpha \mathrm{Pm}^2} - \alpha(\alpha+1) \nonumber \\
 & = & \frac{(\alpha-1)\beta}{\alpha \mathrm{Pm}^2} - \alpha (\alpha+1)
\end{eqnarray}
Hence, the transition is supercritical when $C_3(r = r_c)  < 0$, i.e.,
\begin{equation}
\mathrm{Pm}  > \mathrm{Pm} _c = \frac{1}{\alpha} \sqrt{\frac{(\alpha-1)\beta}{\alpha+1}},
\end{equation}
and subcritical when $\mathrm{Pm}  < \mathrm{Pm} _c$ and $C_5(r = r_c)  <  0$.  It is easy to verify that $C_5(r = r_c)  <  0$ for $\mathrm{Pm}  < \mathrm{Pm} _c$. 

\begin{figure}
\vspace{15pt}
\begin{center}
\includegraphics[scale=0.6]{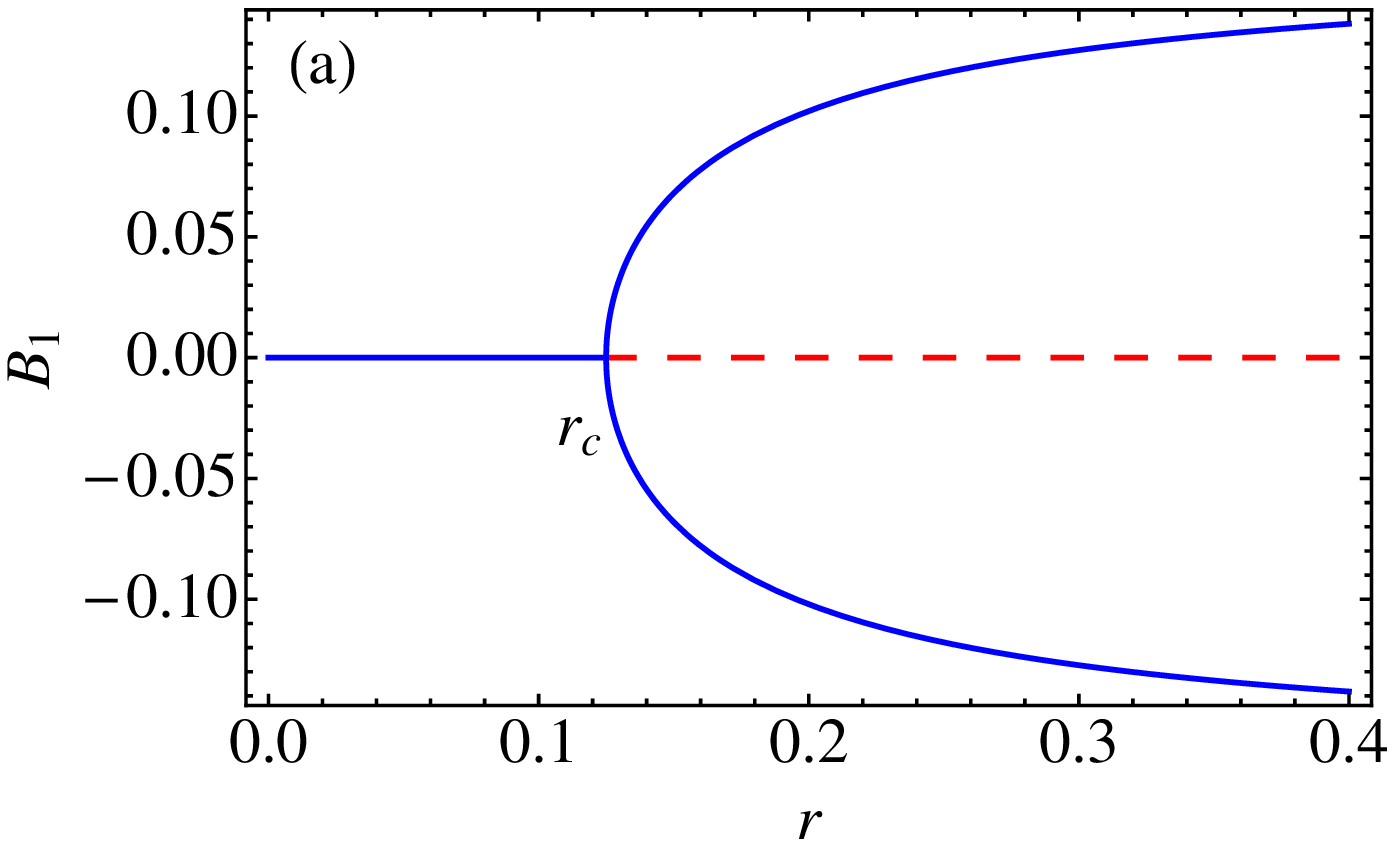} \\
\includegraphics[scale=0.6]{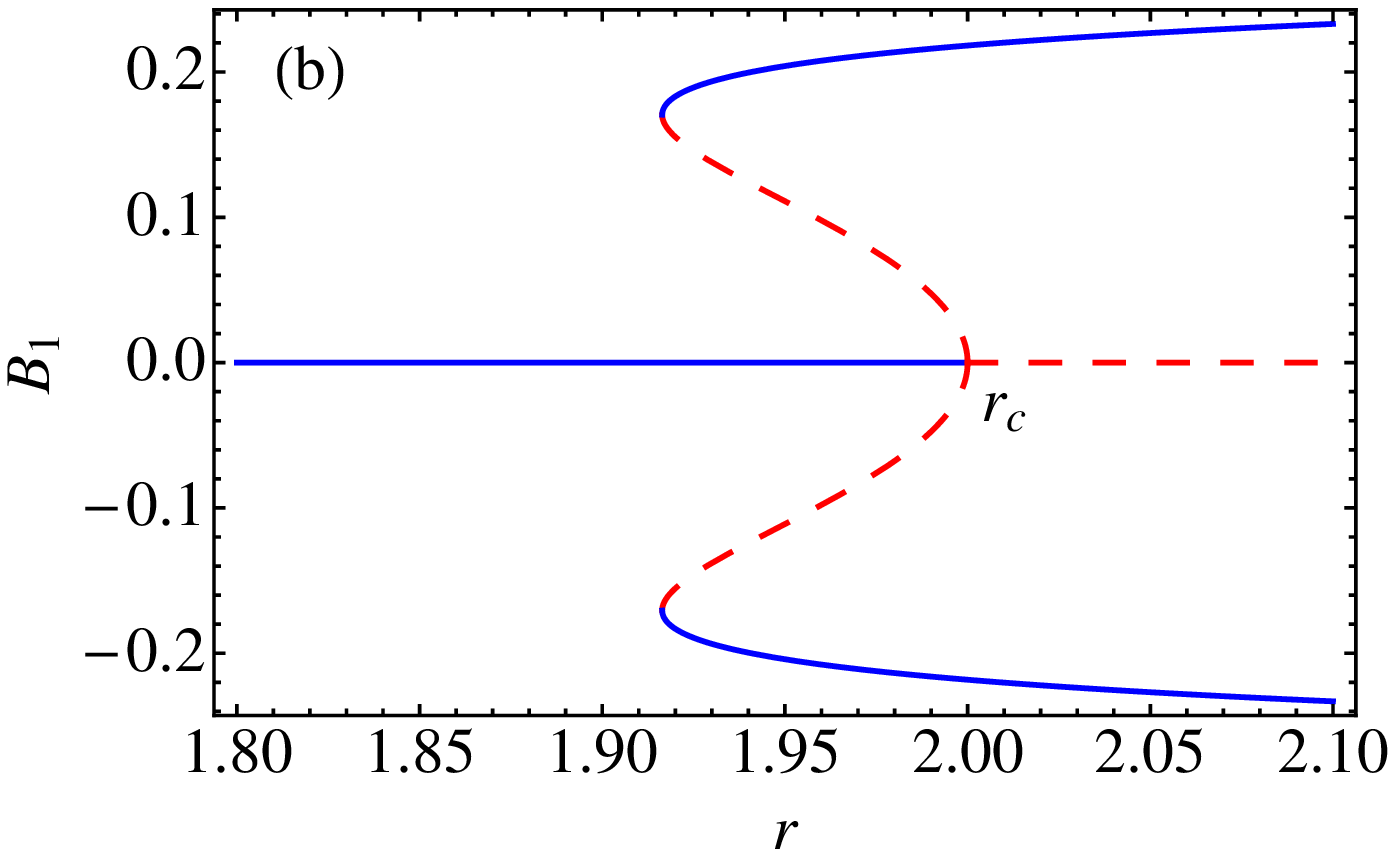}
\caption{Bifurcation diagrams for the 3-mode model of convective dynamo: (a) supercritical dynamo transition for  $\mathrm{Pm}=2$ and (b) subcritical dynamo transition for $\mathrm{Pm}=1/2$.  The transitional Rayleigh number $r_c=1/8$ and 2 respectively for the aforementioned cases.}
\label{fig:convective}
\end{center}
\end{figure}

We illustrate the above transition for the parameter values $\alpha=2$ and $\beta=12$, which yields $\mathrm{Pm}_c=1$.  For these parameters, the dynamo transition is supercritical for $\mathrm{Pm}  > 1$ and subcritical otherwise.  The transition takes place at $r=r_c=1/(2 \mathrm{Pm}^2)$. In Fig.~\ref{fig:convective}, we illustrate the bifurcation diagrams  for $\mathrm{Pm}=2$ (supercritical) and $\mathrm{Pm}=1/2$ (subcritical), for which $r_c=1/8$ and 2 respectively.

The above arguments show that the 3-mode model for the convective dynamo exhibits supercritical and subcritical dynamo transitions  for $\mathrm{Pm} > \mathrm{Pm}_c$ and $\mathrm{Pm} < \mathrm{Pm}_c$ respectively.  These observations are in good qualitative agreement with the numerical simulations of Morin and Dormy~\cite{Morin:IJMPB2009}.   It is also interesting to note that the model does not contain rotation, yet exhibits supercritical and subcritical transitions.  This feature might induce further research to explore the role of rotation, nonlinearity, and convection in dynamo transition.   A deficiency of our model is that it does not capture ``isola" bifurcation~\cite{Morin:IJMPB2009}.  In Appendix~\ref{sec:isola} we briefly discuss a possible extensions of our models that could show Isola bifurcation.

\section{\label{sec:conclusions}Discussions and Conclusions} 
Several numerical simulations indicate that the dynamo transition occurs via supercritical or subcritical bifurcation.  Typically, subcritical bifurcations take place for small magnetic Prandtl number, while the supercritical ones are observed for large magnetic Prandtl numbers~\cite{Yadav:EPL2010, Yadav:PRE2012, Morin:IJMPB2009, Sreenivasan:JFM2011}. The main motivation of our present work is to understand this interesting crossover via low-dimensional models. 

We first establish the nature of a subcritical dynamo transition using direct numerical simulations. We relate a well-known one-dimensional dynamical system (Eq.~(\ref{eq:combined})) to dynamo transition and discuss that the subcritical bifurcations require higher-order nonlinearity as compared to the supercritical ones.  This model also explains why the basin of attraction of the dynamo and the no-dynamo states in the numerical simulations are separated by a well-defined boundary.   

The scope of the one-dimensional model is however limited, and it does not provide us insights into the role of magnetic Prandtl number.  Therefore we construct more sophisticated dynamo models motivated by the symmetries and the nature of the nonlinearities of the MHD equations.    These models contain crucial higher-order nonlinearities of the form $B_1^2 B_2 $ and $B_2^2 B_1 $, which arise due to higher order interactions.   For these models we observe an interesting  crossover from supercriticality to subcriticality as the magnetic Prandtl number is decreased, consistent with the numerical results of the Taylor-Green~\cite{Yadav:EPL2010, Yadav:PRE2012} and spherical-shell dynamos~\cite{Morin:IJMPB2009}.   VKS experiment, which exhibits supercritical dynamo for a very small $\mathrm{Pm}$,  appears to be contrary to the above behavior. The reason for this disagreement is not clearly understood at present.  A strong ``stabilising'' influence of the impellers with high magnetic permeability could be one of the primary reasons~\cite{Giesecke:PRL2010}.   We require further analysis to test this conjecture.

In our 3-mode model, the dynamo transition depends crucially on the extent of stabilizing and destabilizing terms.  Supercritical to subcritical transition takes place when non-linear term similar to the $X^3$ term of Eq.~(\ref{eq:combined}) destabilizes the magnetic field which is subsequently stabilized by an even higher order term.  This feature could be similar to the recent findings of Sreenivasan and Jones~\cite{Sreenivasan:JFM2011} for the subcritical spherical-shell dynamo in which the Lorentz force, which otherwise saturates the magnetic field, destabilizes it.    Interestingly, our model is much simpler than that of  Sreenivasan and Jones~\cite{Sreenivasan:JFM2011}, which includes rotation and complex interplay of helicity.  Future investigation of the role of each part  (e.g., rotation, convection, MHD nonlinearity) would be very valuable.

In summary, our dynamo model exhibits supercritical and subcritical dynamo transitions.  The model also shows a changeover from supercriticality to subcriticality when the magnetic Prandtl number is decreased.   It would be interesting to make quantitative comparisons of the model with realistic dynamos.  However,  the presence of a large number of modes in such systems makes the above task very difficult.   We hope that future works in this direction would provide valuable insights into the dynamo mechanism.


{\bf Acknowledgements:} 
We thank the anonymous referee for very stimulating and helpful comments. We also thank S. Fauve, E. Dormy,  K. P. Rajeev, H. Wanare, and T. Sarkar for useful suggestions and discussions. The numerical simulations were performed on CHAOS cluster (IIT Kanpur) and the VEGA cluster (IIT Madras). The authors also acknowledge the research grant SPO/BRNS/PHY/20090310 and the Swarnajayanti fellowship to M. K. Verma.  


\appendix
\section{\label{sec:nlin} Derivation of the 3-mode model} 

In this appendix, we will derive a 3-mode model using the MHD equations in Fourier space, which are
\begin{eqnarray}
\partial_{t} \hat{ \mathbf{u}}(\mathbf k) + \nu k^2 \hat{ \mathbf{u}}(\mathbf k)  & = & \mathbf{F}(\mathbf k)  -i {\mathbf k} \sigma (\mathbf k) 
 -i \sum_{\mathbf p} \mathbf k \cdot \hat{ \mathbf u}(\mathbf q)\hat{  \mathbf  u}(\mathbf p)  \nonumber \\
& + & i \sum_{\mathbf p} \mathbf k \cdot \hat{ \mathbf b}(\mathbf q) \hat{ \mathbf  b}(\mathbf p)  + (\mathbf p \leftrightarrow \mathbf q) 
\label{eq:MHD_vel_k}\\
\partial_{t} \hat{ \mathbf{b}}(\mathbf k) + \eta k^2 \hat{ \mathbf{b}}(\mathbf k) & = &
- i \sum_{\mathbf p} \mathbf k \cdot \hat{ \mathbf u}(\mathbf q) \hat{ \mathbf  b}(\mathbf p)  \nonumber \\
& + & i \sum_{\mathbf p} \mathbf k \cdot \hat{ \mathbf b}(\mathbf q) \hat{ \mathbf  u}(\mathbf p) + (\mathbf p \leftrightarrow \mathbf q)
\label{eq:MHD_b_k} \\
\mathbf k \cdot \hat{ \mathbf{u}}(\mathbf k) & = & 0,   \\
\mathbf k \cdot \hat{ \mathbf{b}}(\mathbf k) & = & 0,  
\end{eqnarray}
with $\mathbf{k = p + q}$.  The sum is to be performed for all possible triads.   However, in order to derive a low-dimensional model, we perform Galerkin truncation, and focus on a single triad $(\mathbf{k, p,q})$.  We force only the $\mathbf k$ mode, hence $\hat{ \mathbf u}(\mathbf k) $ will be the most dominant mode, and it will induce $ \hat{ \mathbf b}(\mathbf p) $ and $ \hat{ \mathbf b}(\mathbf q) $ magnetic modes.  Therefore, the three modes of our low-dimensional model are $\hat{ \mathbf u}(\mathbf k), \hat{ \mathbf b}(\mathbf p)$, and $ \hat{ \mathbf b}(\mathbf q) $.

We expand the aforementioned velocity and magnetic fields perturbatively as
\begin{eqnarray}
 \hat{ \mathbf u}(\mathbf k)  & = & \hat{ \mathbf u}^{(0)} (\mathbf k) +  \hat{ \mathbf u}^{(1)} (\mathbf k) 
 +  \hat{ \mathbf u}^{(2)} (\mathbf k)  + ...\\
  \hat{ \mathbf b}(\mathbf p)  & = & \hat{ \mathbf b}^{(0)} (\mathbf p) +  \hat{ \mathbf b}^{(1)} (\mathbf p) 
 +   \hat{ \mathbf b}^{(2)} (\mathbf p) + ... \\
   \hat{ \mathbf b}(\mathbf q)  & = & \hat{ \mathbf b}^{(0)} (\mathbf q) +  \hat{ \mathbf b}^{(1)} (\mathbf q) 
 +  \hat{ \mathbf b}^{(2)} (\mathbf q) + ...
\end{eqnarray}
Substitution of the above series in the spectral equations, and collection of terms of the same order yields the following:
\begin{eqnarray}
 \partial_t \hat{ \mathbf u}^{(0)}(\mathbf k)  & = & - \nu k^2 \hat{ \mathbf u}^{(0)} (\mathbf k) +  \hat{ \mathbf F}(\mathbf k)   \label{eq:u0}\\
  \partial_t  \hat{ \mathbf b}^{(1)}(\mathbf p)  & = &  - \eta p^2 \hat{ \mathbf b}^{(1)}  (\mathbf p) - i \mathbf p \cdot \hat{ \mathbf u}^{(0)} (\mathbf k)  \hat{ \mathbf b}^{(1)} (-\mathbf q) \nonumber \\
  & + & i \mathbf p \cdot \hat{ \mathbf b}^{(1)} (-\mathbf q)  \hat{ \mathbf u}^{(0)} (\mathbf k)   \\
   \partial_t \hat{ \mathbf u}^{(2)}(\mathbf k)  & = & - \nu k^2 \hat{ \mathbf u}^{(2)} (\mathbf k)  -i \mathbf k \sigma^{(2)}(\mathbf k) \nonumber \\
   & + & i \mathbf k \cdot \hat{ \mathbf b}^{(1)} (\mathbf p)  \hat{ \mathbf b}^{(1)} (\mathbf q) \nonumber \\
   & + & i \mathbf k \cdot \hat{ \mathbf b}^{(1)} (\mathbf q)  \hat{ \mathbf b}^{(1)} (\mathbf p)    \label{eq:u2} \\
  \partial_t  \hat{ \mathbf b}^{(3)}(\mathbf p)  & = &  - \eta p^2 \hat{ \mathbf b}^{(3)}  (\mathbf p) - i \mathbf p \cdot \hat{ \mathbf u}^{(2)} (\mathbf k)  \hat{ \mathbf b}^{(1)} (-\mathbf q)  \nonumber \\
  & + & i \mathbf p \cdot \hat{ \mathbf b}^{(1)} (-\mathbf q)  \hat{ \mathbf u}^{(2)} (\mathbf k)   \label{eq:b3} \\
    \sigma^{(2)}(\mathbf k) & = & \frac{2}{k^2} (\mathbf k \cdot \hat{ \mathbf b}^{(1)} (\mathbf p))( \mathbf k \cdot \hat{ \mathbf b}^{(1)} (\mathbf q)) \\
 \hat{ \mathbf u}^{(1)}  (\mathbf k)  & = & \hat{ \mathbf b}^{(0)}  (\mathbf p) = \hat{ \mathbf b}^{(2)}  (\mathbf p)  = 0.
\end{eqnarray}
The equation for $\hat{ \mathbf b}^{(1)} (\mathbf q)$ is obtained by interchanging $\mathbf p$ and $\mathbf q$.
Assuming quasistatic approximation for Eq.~(\ref{eq:u2}), i.e., $ \partial_t \hat{ \mathbf u}^{(2)}  (\mathbf k) =  0$, we obtain
\begin{eqnarray}
\partial_t  \hat{\mathbf b}^{(3)} (\mathbf p)   & =  &  - \eta p^2 \hat{ \mathbf b}^{(3)}  (\mathbf p) \nonumber \\
& - & \frac{1}{\nu k^2} (\mathbf p \cdot \hat{ \mathbf b}^{(1)} (-\mathbf q))    (\mathbf k \cdot \hat{ \mathbf b}^{(1)} (\mathbf p))   \hat{ \mathbf b}^{(1)} (\mathbf p) \nonumber \\
& + &  \frac{2}{\nu k^4} (\mathbf k \cdot \hat{ \mathbf b}^{(1)} (\mathbf q))    (\mathbf k \cdot \hat{ \mathbf b}^{(1)} (\mathbf p))    \{ (\mathbf p \cdot \hat{ \mathbf b}^{(1)} (-\mathbf q)) \mathbf k \nonumber \\
& - & (\mathbf p \cdot \mathbf k) \hat{ \mathbf b}^{(1)} (-\mathbf q) \}
\end{eqnarray}
The equation for $\hat{\mathbf b}^{(3)} (\mathbf q) $ can be obtaining by interchanging $\mathbf p$ and $\mathbf q$ in the above equation.   Now a combination of all the above ingredients yields the following  sets of equations for the velocity mode up to second order and the magnetic modes up to the third order:
\begin{gather}
 \partial_t \hat{ \mathbf u}(\mathbf k)   =  - \nu k^2 \hat{ \mathbf u}(\mathbf k) +  \hat{ \mathbf F} (\mathbf k) + i (\mathbf k \cdot \hat{ \mathbf b}  (\mathbf p) ) \hat{ \mathbf b} (\mathbf q)  \nonumber \\
  +  i (\mathbf k \cdot \hat{ \mathbf b}  (\mathbf q) ) \hat{ \mathbf b} (\mathbf p) - i \mathbf k \frac{2}{k^2} (\mathbf k \cdot  \hat{ \mathbf b} (\mathbf p)) (\mathbf k \cdot  \hat{ \mathbf b} (\mathbf q)) \\
 \partial_t  \hat{ \mathbf b}(\mathbf p)   =   - \eta p^2 \hat{ \mathbf b} (\mathbf p) - i \mathbf p \cdot \hat{ \mathbf u} (\mathbf k)  \hat{ \mathbf b}  (\mathbf q) + i \mathbf p \cdot \hat{ \mathbf b} (\mathbf q)  \hat{ \mathbf u}  (\mathbf k) \nonumber \\
 - \frac{1}{\nu k^2} (\mathbf p \cdot \hat{ \mathbf b}(-\mathbf q))    (\mathbf k \cdot \hat{ \mathbf b} (\mathbf p))  \hat{ \mathbf b} (\mathbf p) \nonumber \\
 + \frac{2}{\nu k^4} (\mathbf k \cdot \hat{ \mathbf b}(\mathbf q))    (\mathbf k \cdot \hat{ \mathbf b} (\mathbf p))    \{ (\mathbf p \cdot \hat{ \mathbf b}(-\mathbf q)) \mathbf k - (\mathbf p \cdot \mathbf k) \hat{ \mathbf b}(-\mathbf q) \}  \label{eq:bp_final}
  \end{gather}
The equation for $\partial_t \hat{\mathbf b}(\mathbf q) $ can be derived easily by $\mathbf p \leftrightarrow \mathbf q$ in Eq.~(\ref{eq:bp_final}).
 
The above set of equations are still quite complex. To simplify the algebra and to focus on essential physics of dynamo transition, we make further simplification by  replacing the above  vector equations by scalar equations ($\hat{ \mathbf u}(\mathbf k) = u, \hat{ \mathbf b}(\mathbf p)=b_1, \hat{ \mathbf b}(\mathbf q) = b_2$).   As a result, the appropriately tuned low-dimensional model is
\begin{eqnarray}
\dot{u} & = & f- k_0^2 \nu u- (\alpha+1)k_0 b_{1}b_{2}, \label{eq:u_dot}\\ 
\dot{b}_{1} & = & - k_0^2 \eta b_{1} + \alpha  k_0 u b_{2} - \frac{\beta}{\nu}   b_2^2 b_1,    \label{eq:B1_dot}\\
\dot{b}_{2} & = & - k_0^2  \eta b_{2} + u b_{1} +\frac{\beta }{\nu}  b_1^2 b_2, \label{eq:B2_dot}
\end{eqnarray}
where $k_0 = 2 \pi/L $ with $L$ as the system size. The constants $\alpha,\beta$ are effective constants, and they represent the complex multiplication factors of Eq.~(\ref{eq:bp_final}).  Determination of the value of these constants from the first principle is very involved; here we model them as positive numbers.  In the present paper, our objective is to capture the essential physics of dynamo transition, rather than an accurate representation of TG and convective dynamos. 

We also assume that $k \approx p \approx q \approx k_0$.  Note that the nonlinear terms preserve the $B \rightarrow -B$ symmetry of the MHD equations, as well as conserve the total energy $(u^2+b_1^2+b_2^2)/2$ in the inviscid limit ($\nu=\eta=f=0$).  For $\beta=0$, our model reduces to a form similar to the 3-mode model of Gissinger {\it et al.}~\cite{Gissinger:EPL2010}, which was used for studying  magnetic field reversals in the VKS experiment. 

It is more convenient to use nondimensional equations, which are derived by using $u_0=k_0 \nu$ as the scale for the velocity and magnetic fields, $L$ as the length scale, and $1/(\nu k_0^2)$ as the time scale. That is,  $u=u_0 U$, $b_1 = B_1 u_0$, $b_2 = B_2 u_0$, and $t=T/(\nu k_0^2)$.   Consequently the nondimensionalized equations are
 \begin{eqnarray}
\dot{U} & = & f- U-(\alpha+1)B_{1}B_{2}, \label{eq:App_up_dot}\\ 
\dot{B}_{1} & = & - \frac{1}{\mathrm{Pm}} B_{1} + \alpha U B_{2} -\beta  B_2^2 B_1, \label{eq:App_B1p_dot}\\
\dot{B}_{2} & = & - \frac{1}{\mathrm{Pm}}  B_{2}+U B_{1} +\beta  B_{1}^{2} B_2, \label{eq:App_B2p_dot}
\end{eqnarray}
 
We solve these 3-mode model of TG dynamo in Sec.~\ref{sec:3mode} and explore if the transition is subcritical or supercritical.   We identified the most dominant modes of TG dynamo using our simulation results (Yadav {\it et al.}~\cite{Yadav:EPL2010} and Sec.~\ref{sec:TG_DNS}), according to which $U(2,2,2)$ is the most energetic velocity Fourier mode, and ($B_{1}(0,0,1)$, $B_{3}(-2,-2,1)$)  are the two most energetic magnetic modes.  They correspond to the variables ($U,B_1,B_2$) of our 3-mode model.  Approximate values of the nondimensional kinetic energy and magnetic energy  of some of the dominant modes for $F_{0}=15.3$ are tabulated in Table~\ref{tab:table}.

\begin{table}
\centering
\caption{\label{tab:table} The non-dimensional kinetic energy and magnetic energy of  Fourier modes extracted from the DNS of the TG dynamo for the forcing amplitude of $F_{0}=15.3$ (see Sec.~\ref{sec:TG_DNS}). The most energetic velocity mode is $\hat{\mathbf u}(2,2,2)$, and the two magnetic modes are $\hat{\mathbf b}(0,0,1)$, and $\hat{\mathbf b}(-2,-2,-1)$.}
\begin{tabular}{lcc}
\hline
Fourier mode 	&Kinetic energy &Magnetic energy\\
\hline
$(2,2,2)$	&3.7		&0	\\
$(-2,2,2)$	&0.1		&0	\\
$(-2,-2,0)$	&0.02		&0	\\
$(0,0,1)$	&0		&1.6	\\
$(-2,-2,-1)$	&0		&0.01	\\
$(-2,2,3)$	&0		&0.002	\\
$(-2,-2,1)$	&0		&0.01	\\
\hline
\end{tabular}
\end{table}

\section{\label{sec:isola}Isola in Dynamo Transition and 1D Model} 

\begin{figure}
\begin{center}
\includegraphics[scale=0.45]{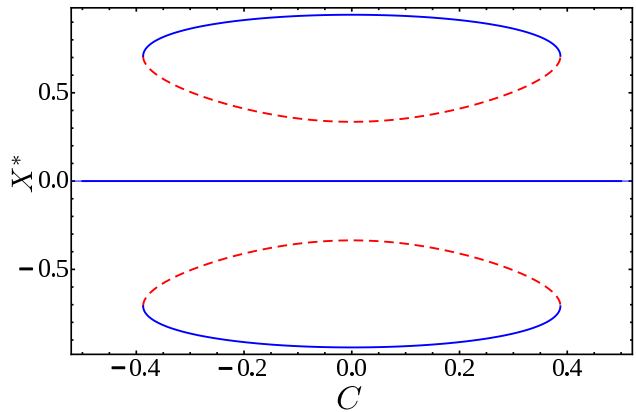}
\caption{Bifurcation diagram for Eq.~(\ref{eq:combined}) exhibiting an Isola transition. Solid (dashed) curves are stable (unstable) branches.}
\label{fig:isola}
\end{center}
\end{figure}

Morin and Dormy~\cite{Morin:IJMPB2009} observed an interesting dynamo transition named ``Isola"  in their spherical dynamo simulations.  In this state, no-dynamo or $B=0$ state is always stable in the parameter regime of the simulation, but a stable dynamo state appears as a part of a ``detached-lobe" for a range of parameter (Rayleigh number).  Interestingly, the Isola can be captured by the one-dimensional model  when $C_3>0$ and $C_5<0$ (similar to the subcritical case), but $C_1$ remaining negative throughout.  To study this case, it is convenient to model 
\begin{equation}
C_1 =  -\xi^2 - C^2,
\end{equation}
with $\xi$ as a constant, and $C$ is a free parameter.   The maximum value $C_1$ can take is $-\xi^2$ when  $C=0$.  Hence we obtain $X^*=0$ as a stable solution for all values of $C$.  In addition we obtain four solutions $X^*_{\pm \pm}$ for $\xi^2 < |C_1| < C_3^2/(4 |C_5|)$ or $|C| \le C_3/\sqrt{4 |C_5|}$.   We illustrate this case in Fig.~\ref{fig:isola} with $\xi^2=0.1$.  Note that $X^*=0, X^*_{+-}, X^*_{--}$ are the stable fixed points, and $X^* = X^*_{++}, X^*_{-+}$ are the unstable fixed points.  These solutions resemble the Isola bifurcations reported by Morin and Dormy~\cite{Morin:IJMPB2009}.

The Isola  is absent in the 3-mode models described in Sec.~\ref{sec:3mode} and \ref{sec:convective_dynamo} since $C_1$ changes sign from negative to positive as the control parameters are increased.  Note that the parameters $\alpha,\beta$ are constants in our model.  However, it may be possible to obtain the Isola if we make the parameter $\alpha$ a nonlinear function of the control parameters such that $C_1<0$ throughout the parameter range.  The exploration of Isola in the TG and convective dynamos is under progress.


\end{document}